\documentstyle[11pt,fleqn]{article}
\begin{document}
\thispagestyle{empty}
\date{}
\title{\bf On the Statistics of CMB Fluctuations Induced by Topological
Defects}
\author{Leandros Perivolaropoulos\thanks{Division of Theoretical Astrophysics,
Harvard-Smithsonian Center for Astrophysics
60 Garden St.
Cambridge, Mass. 02138, USA.}
\thanks{also Visiting Scientist, Department of Physics
Brown University Providence, R.I. 02912, U.S.A.}
}
\maketitle

\begin{abstract}
We use the analytical model recently introduced in Ref. \cite{lp92},
to investigate the statistics of
temperature fluctuations on the cosmic microwave background (CMB), induced by
topological defects.
The cases of cosmic strings and textures are studied. We derive
analytically the characteristic function of the probability distribution for
${\delta T}\over T$
and use it to obtain the lowest twelve moments including the skewness and the
kurtosis. The distribution function is also obtained and it is compared
with the Gaussian distribution thus identifying long non-Gaussian tails. We
show that
for both cosmic strings and textures all odd moments (including skewness)
vanish
while the relative
deviation from the Gaussian for even moments increases with the
order of the moment. The
non-Gaussian signatures of textures, derived from the distribution
function and the moments, are found
to be much more prominent than the corresponding signatures for strings.
We discuss the physical origin of this result.
\end{abstract}

\section{\bf Introduction}

\par
Theoretical models for large scale structure formation can be divided in two
classes according to the type of primordial perturbations they consider;
models based on adiabatic Gaussian perturbations produced during
inflation and models based on non-Gaussian perturbations which are naturally
provided by topological defects. The cold dark matter (CDM) model, based on
Gaussian primordial fluctuations, produces an evolved density field with
small and intermediate scale structure in reasonable agreement with
observations\cite{wdef87}.  Recent observations on large scales however, have
created significant challenges for the CDM model. One of the most serious
such challenges comes from the recent detection of anisotropy in the CMB by
the DMR (Differential Microwave Radiometer) instrument of COBE (Cosmic
Background Explorer) satellite\cite{cb92}.

This discovery
has provided a new powerful experimental probe for testing
theoretical models for large scale structure formation. The temperature sky
maps constructed by DMR were used to obtain the rms sky variation
$\sqrt{<({{\Delta T}\over T})^2>}$ (where $\Delta T\equiv
T(\theta_1)-T(\theta_2)$, and $\theta_1-\theta_2=60^\circ$ is the beam
separation in the COBE experiment) and the rms quadrupole amplitude. A fit of
the data to spherical harmonic expansion has also provided the angular
temperature auto-correlation function $C(\Delta\theta)\equiv  <{{\delta
T}\over T}(\theta)  {{\delta T}\over T} (\theta ^\prime)>$ where $<>$ denotes
averaging over all directions in the sky, $\delta T(\theta)\equiv
T(\theta)-<T>$ and $\Delta \theta=\theta -\theta^\prime$.  This result was
then used to obtain the rms-quadrupole-normalized amplitude
$Q_{rms-PS}$ and the index $n$ of the power law fluctuation spectrum
assumed to be of the form $P(k) \sim k^n$. The published results are:
$$
n=1.1 \pm 0.5
$$
$$
Q_{rms-PS}=(5.96\pm 0.75)\times 10^{-6}
\eqno(1.1)
$$
$$
({{\Delta T}\over T})_{rms}=(1.1 \pm 0.2)\times 10^{-5}
$$
 Severe constraints are imposed on several cosmological models
due to these results. For example, the CDM model with bias
$1.5\leq b_8 \leq 2.5$ is inconsistent with the COBE results for $H_0 > 50
km/(sec\cdot Mpc)$ and is barely consistent for $H_0\simeq 50 km/(sec\cdot
Mpc)$
\cite{cbt92}
\cite{be87}.
It is therefore interesting to investigate the consistency of alternative
models with respect to the COBE measurments. The natural alternative to models
based on adiabatic Gaussian perturbations generated during inflation are models
where the primordial perturbations are created by topological defects like
cosmic strings global monopoles\cite{lp92a} or textures\cite{t89}.

In a recent paper \cite{lp92} we introduced an analytical model and used
it to study the effects of cosmic strings on the microwave background. Our
model
was based on counting random multiple impulses, inflicted on photon
trajectories
by the string network between the time of recombination and today. After
constructing the temperature auto-correlation function, we used it to obtain
the
effective power spectrum index n, the rms-quadrupole-normalized amplitude
$Q_{rms-PS}$ and the  rms temperature variation smoothed on small angular
scales.  For the values of the scaling solution parameters obtained
in Refs.\cite{bb90},\cite{as90} we showed that
$$
n=1.14 \pm 0.5
$$
$$
Q_{rms-PS}=(4.5\pm1.5) G\mu
\eqno(1.2)
$$
$$
({{\Delta T}\over T})_{rms}=5.5 G\mu
$$
where $\mu$ is the mass per unit length of the string (the single free
parameter in the cosmic string model for structure formation) and
$G$ is Newton's constant.

Demanding consistency of our results with the COBE results (1.1) leads to
\cite{lp92}
$$
G\mu=(1.7\pm 0.7)\times 10^{-6}
\eqno(1.3)
$$
in good agreement with direct normalization of $\mu$ from
galaxy\cite{tb86} and large scale
structure\cite{vv91}\cite{pbs90}
observations (for more recent studies of the cosmic string model for
structure formation see Ref. \cite{as92a}). We concluded that the
cosmic string model remains a viable model consistent with the up to now
announced COBE data. Similar results to those presented in Ref. \cite{lp92}
have been obtained by numerical simulations of the string network from the
time of recombination to a small redshift\cite{bbs88}. These
studies however, in contrast to our analytical model, do not attempt to take
into account compensation (for a recent study of the effects of compensation
see Ref.
\cite{vs92}) and allow the string deficit angle to extend over the whole volume
of their
simulation. In addition they are constrained to fixed values of the scaling
solution
parameters produced by simulations which are still subject to some
controversy\cite{at89}\cite{at85}\cite{as90}\cite{bb90}.
Other interesting studies \cite{ttb86} have used the
old picture for the cosmic string network\cite{at85}, based on loops,
to analytically calculate the
effects of strings on the CMB. Recent simulations\cite{bb90}\cite{as90}
however,
have shown that the
dominant component of the scaling solution consists of long strings rather than
loops.

Our analysis had utilized the data concerning the {\it amplitude} and the
{\it spectrum} of the detected fluctuations in order to test the cosmic string
model. This type of test can check the consistency of the model with the data
but it can not distinguish it from other consistent theories.
There are two basic reasons for this:
First, the $n=1$ Zeldovich spectrum is fairly generic in physically motivated
theories
and second, other theories like standard CDM, can also pass the
amplitude normalization test
\cite{s92b} by utilizing tensor mode
perturbations.

It is therefore clear that further tests are needed in order to distinguish
between the topological defect models and other theories. One of the most
interesting such tests is the subject of this paper:
{\it the study of the statistics of the CMB fluctuations}.
This is a particularly interesting issue
in view of the existing temperature fluctuation sky maps obtained by COBE
which are currently subjected to statistical analysis by the COBE
collaboration. Such an analysis can reveal the characteristic signature of
the source that produced the CMB fluctuations. The identification of this
signature for topological defect models is the focus of the present work.

Models based on fluctuations generated during inflation predict (in their
generic form) a Gaussian distribution function for the  CMB temperature
fluctuations. On the other hand, models based on topological defects, like
cosmic strings or textures, are distinguished due to the particular type of
non-Gaussian fluctuations they predict.

There is an extensive literature on the statistical effects of various types
of non-Gaussian perturbations (for a small sample see
Ref. \cite{ls92}).  On the other
hand the literature on the statistics of seed-like perturbations is much more
limited (see e.g. \cite{gpjbbbs90}\cite{b92}\cite{sb91}).  Previous analytical
studies on the type of non-Gaussian statistics induced by cosmic strings and
other seed-like perturbations have focused on the effects of  point-like
seeds on the statistics of large scale structure\cite{sb91}. They have
considered the superposition of overdense, spherically symmetric kernels,
thus obtaining the density distribution function and other statistical
properties. Those studies, even though elegant and convienient for large
scale structure considerations can not be easily applied to the CMB case. The
basic reason for this, is the need for superposition of variable size (and in
general non-spherical) kernels to account for the effects of compensation in
different Hubble times and horizon scales. The multiple impulse approximation
of Ref. \cite{lp92} (see also Ref. \cite{v92} for an alternative application
of the method) can incorporate this kernel variability and it is the method
that we used in this work.

In particular, we use the multiple impulse approximation to derive
the characteristic function, the temperature distribution function and several
moments of cosmic string and texture induced CMB fluctuations. These results
are then compared with the corresponding Gaussian results in an effort to find
signatures of the cosmic string and texture models.
In section 2 we give a brief review
of our model in order to clarify the basic assumptions made (for a more
complete
account see Ref.\cite{lp92}). We also associate with it a set of
statistical experiments. In section 3 we introduce some basic statistical
quantities and derive the characteristic functions for
the statistical processes described in section 2. In particular we derive the
characteristic functions that correspond to string, texture and Gaussian
temperature paterns. Finally, in section 4 we use the characteristic
functions to obtain the probability distributions and the lowest twelve
moments in each of the three cases. We also compare and discuss the obtained
results.

\section{\bf The Model}

We begin by reviewing the model introduced in Ref. \cite{lp92} which was
originally designed to approximate the cosmic string induced temperature
fluctuations.
It will be seen that with some modifications it can also describe
texture induced temperature fluctuations.

According to cosmic string simulations\cite{at89}\cite{bb90}\cite{as90}, at any
time
t there are about 10 long string  segments with a typical curvature radius t
passing through each horizon volume. There is a well defined
mechanism\cite{ks84}
by which these segments give rise to localized linear temperature
discontinuities
(Fig. 1). Photons passing on different sides of a long straight string moving
with
velocity $v_s$ perpendicular to the line of sight, reach the observer with a
Dobbler
shift\cite{ks84}
$$
{{\delta T}\over T}   = \pm 4\pi G\mu v_s \gamma_s
$$
Our goal is to find the combined effects of temperature fluctuations induced by
strings present between the time of recombination $t_{rec}$ and today.

 We approximate the photon path from the recombination
time $t_{rec}$ to the present time $t_0$ by a discrete set of $M=16$ Hubble
time-steps $t_i$ such that $t_{i+1}=2 t_i$ (${t_0\over t_{rec}}\simeq 2^{16}$
for
$z_{rec}=1400$). We assume that the effects of long strings between the time of
recombination and today give the dominant contribution to temperature
fluctuations
and therefore  we consider photons emerging from the last scaterring surface at
$t_{rec}$ with a fixed uniform temperature. A beam of photons coming from a
fixed
direction will initially suffer $n_h$ `kicks' from the $n_h\simeq 10$ long
strings
within the horizon at $t_1 = t_{rec}$ (the linear superposition of `kicks' is
consistent with the multi-string metric presented in Ref.\cite{g91}).
Each temperature `kick' from a string with arbitrary orientation with
respect to the observer induces a fluctuation ${\delta T}\over T$ of the form
$$
{{\delta T}\over T}   = \pm 4\pi G\mu \beta
\eqno(2.1)
$$
with
$$
\beta={\hat k}\cdot({\vec v_s}\gamma_s \times {\hat s})
$$
where $\hat{k}$, $\hat{s}$ and $\vec{v}_s$ are the unit wave-vector, the unit
vector along the string and the string velocity vector respectively.
The sign changes along the string\cite{ks84}.
%
%\footnote{ Notice that the $\pm$ sign in (1.3) is
%important when studying statistics but can be omitted without significant
% error
%when looking at average values of correlated `kicks' as was done in paper I.}.
%In this paper we take $\beta$ to be a discrete random variable with uniform
%distribution.
%
At the next Hubble time-step $t_2$, $n_h$ further `kicks',
uncorrelated with the  initial ones, will be inflicted on the photon beam by
the
strings present within the new horizon scale $t_2 = 2 t_1$. The process will
continue until the $M=16$ Hubble time-step corresponding to the present time
$t_0$ (Fig. 2).

The physical process described above, corresponds to the following set of
successive statistical experiments. For simplicity we will first consider the
one dimensional case. The generalization to the realistic two dimensional case
will then be made in a straightforward way.

 Consider a one dimensional
(continuous or not) set of temperature pixels with initially uniform
temperature distribution and fixed total length L.  Consider now a
localized step-function perturbation imposed on this surface. Such a
perturbation would increase the temperature of a fixed small length $l_0$ by an
amount ${{\delta T}\over T}=\delta$ but would decrease the temperature of a
neighbouring equal length by the same amount (see Fig. 3). Thus after this
`trial' each pixel of the set has probability $p(1)={l_0\over L}\equiv p_0$ to
have
been positively perturbed, the same probability to have been negatively
perturbed ($p(2)=p(1)\equiv p_0$) and probability $p(3)=1-{2l_0\over L}=1-2
p_0$ to
have remained unperturbed. Let this `trial' repeat $n_0$ times before the first
`experiment' is completed. The next step is to repeat this experiment with a
new
scale for the step-function $l_1=2 l_0$ and $n_1=n_0/2$ number of `trials'. The
successive experiments continue until $l_M=l_0 2^M$ and $n_M=n_0/2^M$. We
demand
$l_M={L\over 2}$ and $n_M=n_h$ where $n_h$ is a fixed positive integer to be
identified  with the number of long strings per horizon volume (see below).
Therefore $l_0={L\over 2^{M+1}}$ and $n_0=n_h 2^M$. This implies that
$$
p_j= 2^j p_0={2^j\over 2^{M+1}}
\eqno(2.2)
$$
$$
n_j={n_0 \over 2^j}= n_h {2^M \over 2^j}
\eqno(2.3)
$$
with $j=0,1,...,M$.
The above described statistical process corresponds to our physical model
provided that the following identifications are made:
\begin{itemize}
\item
The fixed length L is identified with the scale of the present horizon.
\item
Each step-function perturbation with scale $l_j$ is identified with a
cosmic string perturbation induced at the Hubble time-step $t_j$. At
this time the horizon scale is $t_j=2 l_j$. Compensation confines the
string induced perturbation within the horizon scale $2 l_j$. Clearly $2
l_0$ is to be identified with the horizon scale at recombination while $t_M=2
l_M$ (M is the final step) should be identified with the present horizon
scale L.
\item
Each `experiment' is identified with a Hubble time-step. During
the $j$th `experiment' there are $n_j={{n_h 2^M}\over 2^j}$ `trials' (string
perturbations) corresponding to the number of strings per horizon length
(volume) $n_h$ times the total number of horizons $2^M\over 2^j$
within the fixed length L
at the $j$th step. Clearly for $j=M$ there is only the present
horizon included and the total number of `trials'(string perturbations) is
$n_M=n_h$.
\end{itemize}
The above described process needs to be improved in two ways in order to
correctly approximate the physics:
\begin{enumerate}
\item
The one dimensional set of pixels must be promoted to a two dimensional one.
\item
The amplitude of the step-function perturbation must be allowed to vary in
order to account for varying string velocities (varying parameter $\beta$ in
equation (2.1)). Without introducing this improvement our analysis would still
be interesting but it would approximate the perturbations induced by
textures rather than strings. In the case of textures there is no velocity
parameter involved but there are still equal magnitude positive and negative
fluctuations depending on whether the photon `falls in' or `climbs out of' the
texture\cite{ts90}.
\end{enumerate}
Before we incorporate these modifications we proceed to study the statistics of
the perturbations in the above described simplest case. This will help
illustrate our method more clearly while making its generalization a simple and
straightforward task.

\section{\bf The Statistics}

Let us first focus on a particular `experiment' $j$ (Hubble time-step).
Define $f(n_j,k_j)$ to be the probability that any
given pixel will have been perturbed by ${{\delta T}\over T}=k_j \delta$ at the
end of the $n_j$ `trials' (string perturbations).
 Since at each `trial' there are three possible
outcomes with known fixed probabilities $p(1)=p(2)={l_0\over L}\equiv p_j$,
$p(3)=1-2{l_j\over L}=1-2p_j$, the probability distribution $f(n_j,k_j)$ may
be obtained from the well known trinomial distribution (a simple
generalization of the binomial). The trinomial distribution $f_{n_j}
(x(1),x(2),x(3))$ gives the probability that after $n_j$ `trials' with three
possible outcomes, there are $x(i)$ occurences of outcome $i$
($i=1,2,3$) (obviously $x(1)+x(2)+x(3)=n_j$). The trinomial
distribution is:   $$
f_{n_j} (x(1),x(2),x(3))={{n_j!}\over {x(1) ! x(2) ! x(3) !}}
p(1)^{x(1)} p(2)^{x(2)} p(3)^{x(3)}
\eqno(3.1)
$$
Let $x(1)$ ($x(2)$) be the number of positive (negative) temperature shifts for
a given pixel while $x(3)$ is the number of `trials' that lead to no shift.
Using the relations $x(1)+x(2)+x(3)=n_j$ and $k_j=x(1)-x(2)$  to change
variables
from $x(1),x(2),x(3)$ to $n_j,k_j,x(3)$ and summing over the possible $x(3)$ we
obtain
$f(n_j,k_j)$:
$$
f(n_j,k_j)=\sum_{x(3),2}{{n_j! p_j^{n_j-x(3)} (1-2p_j)^{x(3)}}\over
{({{n_j+k_j-x(3)}\over 2})!  ({{n_j-k_j-x(3)}\over 2})! {x(3)}!}}
\eqno(3.2)
$$
where the sum extends over all integer values of $x(3)$ for which $n_j\pm
k_j-x(3)$ is even. (notice that terms involving factorials of
negative numbers vanish automatically).

A more useful and much simpler function that describes the statistics is the
{\it characteristic function} $\phi (n,\omega)$ of the distribution. This has
two
important properties:
\begin{enumerate}
\item
It is the Fourier transform of the probability distribution i.e.
$$
\phi (n,\omega)=\sum_{k=-n}^n e^{i\omega k} f(n,k)
\eqno(3.3)
$$
$$
f(n,k)={1\over {2 \pi}}\int_0 ^{2\pi} e^{-i \omega k} \phi (n,\omega)
\eqno(3.4)
$$
\item
It can generate all moments $<k^m>$ of the distribution by differentiation i.e.
$$
<k^m>=(-1)^m i^m {{d^m} \over {d \omega ^m}}\phi (n,\omega)\vert_{\omega = 0}
\eqno(3.5)
$$
\end{enumerate}

Using the property
$$
(p(1)+p(2)+p(3))^n=\sum_{x(1)+x(2)+x(3)=n} f_n(x(1),x(2),x(3))
$$
and equation (3.3) it is straightforward to show that the characteristic
function for the variable $k_j=x(1)-x(2)$ is:
$$
\phi (n_j,p_j,\omega)=(2p_j\cos(\omega) + (1-2p_j))^{n_j}
\eqno(3.6)
$$
However, we are interested in a multiple `experiment' process i.e. we are
looking for the distribution function of the variable:
$$
K=\sum_{i=0}^{M\simeq 16} k_i
\eqno(3.7)
$$
where $K$ is to be identified with the total temperature fluctuation ${{\delta
T}\over T}$ at the present time $t_0$. It is straightforward to show \cite{f71}
that
the characteristic function for a sum of independent random variables is equal
to
the product of the individual characteristic functions. Therefore, the
characteristic function $\Phi(\omega)$ corresponding to the variable K is:
$$
\Phi (\omega)= \prod_{j=0}^{M\simeq 16} \phi(n_j,p_j,\omega)
\eqno(3.8)
$$
This result may now be used to obtain the probability distribution by the
Fourier transform (3.4). It may also be used to obtain all the moments of the
distribution either by differentiation (using equation (3.5)) or by direct
integration, using the distribution function. However, we must first identify
correctly the parameters $n_j$, $p_j$ for the physical problem under
consideration. Clearly, equations (2.2), (2.3) need to be modified to account
for
the propagation of a surface (photon wavefront) rather than a line, through the
string network. It is straightforward to generalize arguments  leading to (2.2)
and (2.3) to the two dimensional case to obtain:
$$
p_j= 4^j p_0={2\times 4^j\over 4^{M+1}}
\eqno(3.9)
$$
$$
n_j={n_0 \over 4^j}= n_h {4^M \over 4^j}
\eqno(3.10)
$$
Using (3.9) and (3.10) in (3.8) we obtain:
$$
\Phi_t (\omega)= \prod_{j=0}^{M\simeq 16} (4^{j-M} \cos(\omega) + (1-
4^{j-M}))^{n_h 4^{(M-j)}}
\eqno(3.11)
$$
where the subscript $t$ denotes that this result is valid only for the case of
texture-like perturbations where the magnitude of the individual perturbations
(step-function magnitude) can be considered fixed. In the case of strings we
must generalize (3.11) further in order to account for the variable parameter
$\beta$ of equation (2.1) which represents the velocity and string orientation
dependence of the perturbations. This generalization will lead to the
multinomial distribution.

Consider the above described process of successive perturbations with the
additional feature of allowing $Q$ possible magnitudes for the applied
step-function perturbations. Let $p_j^i$ be the probability for any
temperature pixel to be perturbed $i$ units at the $j$th Hubble time-step where
$i=-Q,...,+Q$ and $j=0,...,M$. We now have a total of $2Q+1$ possible outcomes
for each `trial'. Therefore, the distribution function can be obtained from a
generalization of the trinomial:
{\it the multinomial distribution}. The
multinomial distribution $f_n (x(1),...,x(R))$ gives the probability that an
experiment consisting of n `trials' each with R possible outcomes will result
to
$x(i)$ occurrences of the $i$th outcome ($i=1,...,R$) given that the
probabilities for each outcome are $p(1),...,p(R)$. In direct correspondence
with the trinomial, $f_n (x(1),...,x(R))$ is the general term of the
multinomial
expansion $(p(1)+p(2)+...+p(R))^n$. The interesting variable for our purposes
is
the total temperature fluctuation which at the $j$th time-step is given by
$$
k_j=\sum_{i=1}^Q i (x_j^i-x_j^{-i})
$$
where $x_j^i$ is the number of $i$ unit perturbations at the $j$th Hubble
time-step.
Using the
multinomial expansion and equation (3.3), it is straightforward to obtain the
characteristic function
for the distribution of $k_j$. The result is:
$$
\phi (n_j,p_j^1,...,p_j^Q,\omega)=(2\sum_{i=1}^Q p_j^i \cos(i\hspace{2mm}\omega
) +
(1-2 \sum_{i=1}^Q p_j^i)^{n_j}
\eqno(3.12)
$$
In order to proceed further we must specify the probability distribution of
the step-function magnitudes i.e. the dependence of $p_j^i$ on the index $i$.
The simplest physically interesting choice is the distribution
$$
p_j^i={{2\times 4^{j-M-1}}\over Q}
\eqno(3.13)
$$
which corresponds to a uniform distribution of the parameter $\beta$ in
equation (2.1) in the range $[0,\beta_{max}]$ where $\beta_{max}$ should be
chosen to be about unity.

 Using now (3.12), (3.13) and (3.10) in (3.8) we obtain the generalization of
(3.11)
appropriate for cosmic strings:
$$
\Phi_s (\omega)= \prod_{j=0}^{M\simeq 16} ({4^{j-M}\over Q} \sum_{i=1}^Q
\cos(i\hspace{2mm}\omega) + (1- 4^{j-M}))^{n_h 4^{M-j}}
\eqno(3.14)
$$
which along with (3.11) is the central result of this work. In what follows we
use the results (3.11) and (3.14) to obtain the probability distributions and
several moments for temperature fluctuations induced by textures and cosmic
strings. In accordance with the accepted values of the scaling solution
parameters we will use $n_h=10$ for strings\cite{bb90}\cite{as90} and
$n_h=0.04$ for
textures\cite{stpr91}. Our goal is to compare the derived results with those
corresponding
to a Gaussian distribution. It is therefore instructive to first investigate if
and under
what conditions can we obtain a Gaussian limit for the characteristic functions
(3.11) and
(3.14).

Consider first the texture case described by equation (3.11).
Since we want to compare with the standard Gaussian distribution which has
$\sigma=1$ we must first appropriately normalize the variable $K$  dividing by
its
standard deviation to match the standard deviation $\sigma=1$ of the standard
Gaussian.
Consider the new variable
$$
K_g=\sum_{i=0}^M {{k_i}\over \sqrt{{n_h (M+1)}}}
\eqno(3.15)
$$
The characteristic
 function that  generates the moments of $K_g$ is
$$
\Phi_t^g(\omega)=\Phi_t({{\omega}\over {\sqrt{n_h (M+1)}}})
\eqno(3.16)
$$
It is now straightforward to show that
$$
\lim_{n_h\to \infty}{\Phi_t^g (\omega)}=\lim_{n_h\to \infty}{\prod_{j=0}^M
[(1-{{(4^{j/2}\omega
/\sqrt{M+1})^2} \over{ 2n_h 4^M}})^{n_h 4^M}]^{1/4^j}}=e^{-\omega^2/2}
\eqno(3.17)
$$
But $e^{-\omega^2/2}$ is the characteristic function for the standard
Gaussian\cite{f71} distribution.
Therefore the distribution of the appropriately normalized variable $K_g$
approaches the standard Gaussian for $n_h\longrightarrow \infty$. Notice that
the
Gaussian limit is not obtained for $M \longrightarrow \infty$ i.e. for the
Gaussian limit to be realized we need several perturbations per horizon volume
but evolution over more Hubble time-steps does not help.
In a similar way, it may be shown that the Gaussian limit for the string
characteristic function (equation (3.14)) is obtained for the variable
$$
K_g=\sum_{i=0}^M {{k_i}\over \sqrt{{n_h (M+1)(Q+1)(2 Q+1)/6}}}
\eqno(3.18)
$$
provided that $n_h \longrightarrow \infty$. For this variable the string
characteristic function is:
$$
\Phi_s^g(\omega)=\Phi_s({{\omega}\over {\sqrt{n_h (M+1)(Q+1)(2 Q+1)/6}}})
\eqno(3.19)
$$

\section{Results-Discussion}

We are now in position to use (3.11) and (3.14) (or (3.16) and (3.19))
to make predictions and compare
these with the predictions of the corresponding Gaussian distribution.
We first obtain the moments of the distributions. This may
be achieved by differentiating the characteristic functions and using
equation (3.5) to obtain the moments $<K^m>$. Alternatively we may expand the
characteristic functions around $\omega=0$. It is the later method we used
here. We  used the symbolic manipulation package {\it Mathematica}\cite{w92}
to expand (3.16) and (3.19)  around $\omega =  0$ up to order 12. From the
values of the coefficients and equation (3.5) we obtained the 12 lowest moments
for
the texture and string cases and also for the standardized Gaussian.
The values of these moments are shown in Table 1.
\vskip 0.5cm
{\bf Table 1}:The values of the lowest six non-vanishing moments for the string
($<K_g^m>_s$), texture ($<K_g^m>_t$) and standard Gaussian ($<K_g^m>_G$).
\vskip 0.1cm
\begin{tabular}{|c|c|c|c|}\hline
$m$ & $<K_g^m>_s$ & $<K_g^m>_t$ & $<K_g^m>_G$ \\ \hline
2 & 1    & 1 & 1 \\ \hline
4 & 3.0045 & 4.124 & 3 \\ \hline
6 & 15.0675 & 35.5577 & 15 \\ \hline
8 &  105.947 & 440.226 & 105 \\ \hline
10 & 959.239 & 9681.3 & 945 \\ \hline
12 &  10630 & 159955 & 10395 \\ \hline
\end{tabular}
\vskip 0.5cm

The
quantity of interest is the relative deviation from the Gaussian defined as:
$$
R_m^{s,t}={{<K_g^m>_{s,t}-<K^m>_G}\over <K^m>_G}
\eqno(4.1)
$$
where $<K_g^m>_{s,t}$ is the standardized string or texture $m$th moment while
$<K^m>_G$ is the
corresponding standard Gaussian moment.

In Fig. 4 we plot the relative deviation $R_m^s$ versus $m$ for strings (notice
that only even moments are shown since all odd moments vanish) obtained
by expanding  (3.19) with $M=16$, $Q=5$ and $n_h=10$.  Clearly, $R_m^s$ is a
rapidly increasing
function of $m$ which is evidence for the presence of long non-Gaussian tails
in the
string distribution function. However, even for the $m=12$ moment the relative
deviation does not exceed 3\%  implying that the non-Gaussian features in the
string
distribution function are fairly weak.
 The reason for this is that in the case of strings we have $n_h=10 \gg 1$
which implies that the Gaussian limit is approached in an effective way. We
have
performed tests for different values of Q and found that the values of
$R_m^s$ remain insensitive to within a factor of 2.

In Fig. 5 we plot the corresponding relative deviations $R_m^t$ for textures.
 We used equation (3.16) with $M=16$ and
$n_h=0.04$. In the case of textures, not only is $R_m^t$ a rapidly increasing
function of $m$ but also even the lower moments  are significantly larger than
the
corresponding Gaussian. For example the kurtosis (defined as
$<K_g^4>/({{<K_g^2>}^2})$) is predicted to be 40\% larger than the kurtosis
of the standard Gaussian distribution, while
the sixth moment is larger by  a factor of three.  As in the case of strings,
the
skewness and all the odd moments are found to vanish. This is due to the fact
that
the superimposed kernels are symmetric with respect to positive and negative
perturbations. Clearly, such an assumption even though reasonable for CMB
consideration is inapplicable for large scale structure calculations where a
non-zero
skewness is predicted by seed-based models\cite{sb91}\cite{ls92}.

The characteristic
functions (3.11) and (3.14) can also be used to find the temperature
fluctuation
distribution functions. These can be obtained by Fourier transforming the
characteristic functions according to equation (3.4). The results may then be
compared with the corresponding Gaussian with the same standard deviation (in
order
to keep the Fourier transform simple, we use the original forms (3.11) and
(3.14) in this
case). In Fig. 6a we show the distribution function $F_s (K)$ for strings,
obtained by Fourier
transforming (3.14) with $M=16$, $Q=5$ and $n_h=10$. The difference $F_s(K)-F_G
(K)$ between
the string induced distribution function $F_s (K)$ and the corresponding
Gaussian $F_G (K)$
is shown in Fig. 6b. The relative difference $(F_s(K)-F_G (K))/F_G (K)$ at any
given point
does not exceed 1\% but the presence of long non-Gaussian tails is
clear.  It is these tails that cause the rapid increase of the moments with the
order $m$.

Fig. 7a shows the distribution function for textures obtained by Fourier
transforming (3.11) with $M=16$ and $n_h=0.04$. Superimposed is the
corresponding
Gaussian distribution function.
In this case, as expected since $n_h \ll 1$, the non-Gaussian features
are fairly clear. The central pronounced peak and the long tails seem to be
generic features for temperature fluctuations induced by topological defects.
Fig. 7b shows the difference $F_t(K)-F_G (K)$ between the texture distribution
and the corresponding Gaussian. It shows the same features as Fig. 6b which
applies to
strings but in the case of textures the magnitude of the relative difference
$(F_t(K)-F_G(K))/F_G (K)$ is almost two orders of magnitude larger.
This is reflected in
the relative deviation of moments for which we have $R_m^t \gg R_m^s$.

 Our results on both the relative deviation of moments from the
Gaussian and the distribution function itself show the following:
\begin{enumerate}
\item
The non-Gaussian signature of cosmic strings is difficult to detect by
measuring relative deviation of moments from the Gaussian. Relative deviations
need to be measured to within less than 1\% in order to distinguish
cosmic strings fluctuations from Gaussian. The origin of this approximatelly
Gaussian behavior of string induced perturbations is the large number of
strings per horizon volume ($n_h\simeq 10$). The large number of
superimposed non-Gaussian features `averages out' to an approximatelly Gaussian
result as predicted by the central limit theorem.
\item
The measurment of moments provides a much more powerfull test for the texture
model. This is due to the small number of textures unwinding per horizon
volume ($n_h\simeq 0.04$) which avoids the suppression of the texture
non-Gaussian features. Measuring the relative deviation of the kurtosis to
within 40\% should be enough to detect the deviation induced by textures. For
the relative deviation of the sixth moment, a measurment
accurate to within a factor of less than three, would be sufficient to indicate
the
presence of a texture signature.
\end{enumerate}

The tests based on relative deviation of moments from the Gaussian that have
been
studied here, have several interesting and powerful features, particularly for
testing
the texture model. However, they are not sensitive to geometrical and
topological
features of the temperature fluctuation maps. Such features have been examined
in Ref. \cite{gpjbbbs90} using a numerically obtained realization of cosmic
string induced perturbations. It was found that topological and geometrical
tests can be a sensitive probe of stringy non-Gaussian features.
An interesting extension of the work presented here is the study of the
geometrical
features of string and texture induced temperature patterns using
analytical methods and Monte Carlo simulations. Such a project is currently in
progress\cite{bmp92}.
\section {Acknowledgements}
\par
I wish to thank R. Brandenberger and R. Moessner for
interesting discussions and for providing helpful comments after reading the
paper.
This work was supported by a CfA Postdoctoral Fellowship.

\section{Figure Captions}

{\large \bf Figure 1:} The production of step-like discontinuities in the
microwave temperature for photons passing on different sides of a cosmic string
S.
The string deficit angle is $\alpha$ and O is the observer.
\vskip .5cm
{\large \bf Figure 2:} The propagation of a photon beam from the recombination
time $t_{rec}$ to the present time $t_0$. The horizon in three successive
Hubble
timesteps is also shown.
\vskip .5cm
{\large \bf Figure 3:} The effects of a step-function perturbation on an
initially uniform one dimensional distribution.
\vskip .5cm
{\large \bf Figure 4:} The relative deviation of moments from
the standard Gaussian. $R_m^s$   corresponds to moments due to string induced
perturbations and is plotted versus the $m$ where $m$ is the order of the
moments.
Odd moments are omitted since they vanish.
\vskip .5cm
{\large \bf Figure 5:} The relative deviation $R_m^t$ for the case of textures.
The
deviations from the Gaussian are much larger compared to the case of strings.
\vskip .5cm
{\large \bf Figure 6a:} The distribution function $F_s(K)$ for string induced
perturbations. \\
{\large \bf Figure 6b:} The difference $F_s (k) - F_G (K)$ where $F_G (K)$ is
the
Gaussian distribution with the same standard deviation as $F_s (K)$. The {\it
relative} difference does not exceed 1\% but the presence of long non-Gaussian
tails
is clear.
\vskip .5cm
{\large \bf Figure 7a:} The distribution function $F_t(K)$ for texture induced
perturbations superimposed with the Gaussian distribution of the same standard
deviation.\\
{\large \bf Figure 7b:} The difference $F_t (k) - F_G (K)$ where $F_G (K)$
is the Gaussian distribution with the same standard deviation as $F_t (K)$. The
{\it
relative} difference exceeds 10\% and is much more prominent than in the case
of
strings.

\end{document}